\documentclass[12pt]{article}
\usepackage{amsmath, setspace,geometry,color}
\usepackage{graphicx}
\usepackage{lineno}
\usepackage{lscape}

 \makeatletter       
 \renewcommand{\@biblabel}[1]{#1.}
 \makeatother
\begin{document}
\noindent
{\LARGE\textbf{Classification of volcanic ash particles using a convolutional neural network and probability}} \\
\\
\\
{\large Daigo Shoji$^{1*}$, Rina Noguchi$^{2}$, Shizuka Otsuki$^{3}$, Hideitsu Hino$^{4}$}\\
\\
1. \textit{Earth-Life Science Institute, Tokyo Institute of Technology, 2-12-1 Ookayama, Meguro-ku, Tokyo}\\
2. \textit{Volcanic Fluid Research Center, School of Science, Tokyo Institute of Technology, 2-12-1, Ookayama, Meguro-ku, Tokyo}\\
3.\textit{Geological Survey of Japan, AIST, 1-1-1 Higashi, Tsukuba, Ibaraki}\\
4.\textit{The Institute of Statistical Mathematics,10-3 Midori-cho, Tachikawa, Tokyo}\\
$*$ Corresponding author: shoji@elsi.jp
\newpage
\section*{Abstract}
\textbf{Analyses of volcanic ash are typically performed either by qualitatively classifying ash particles by eye or by quantitatively parameterizing its shape and texture. While complex shapes can be classified through qualitative analyses, the results are subjective due to the difficulty of categorizing complex shapes into a single class. Although quantitative analyses are objective, selection of shape parameters is required. Here, we applied a convolutional neural network (CNN) for the classification of volcanic ash. First, we defined four basal particle shapes (blocky, vesicular, elongated, rounded) generated by different eruption mechanisms (e.g., brittle fragmentation), and then trained the CNN using particles composed of only one basal shape. The CNN could recognize the basal shapes with over 90\% accuracy.  Using the trained network, we classified ash particles composed of multiple basal shapes based on the output of the network, which can be interpreted as a mixing ratio of the four basal shapes. Clustering of samples by the averaged probabilities and the intensity is consistent with the eruption type. The mixing ratio output by the CNN can be used to quantitatively classify complex shapes in nature without categorizing forcibly and without the need for shape parameters, which may lead to a new taxonomy.} 

\section*{Introduction}
The shape of volcanic ash particles depends on the eruption type. Therefore, by analyzing ash shape, we can better understand the physical mechanism of volcanic eruptions, which can help to mitigate volcanic hazards. Classification of volcanic ash particles has traditionally been performed visually$^{1-3}$. Although humans can recognize and categorize complex shapes qualitatively without carrying out calculations, due to biases and differences in level of experience, the result of such classification is often subjective. When we classify objects, they are typically categorized into a single class that is independent from other classes. However, the shapes of natural objects such as volcanic ash are often too complex to categorize into a single class, even for experienced people. 

In order to solve this problem, objective methods using shape parameters have been used$^{4-14}$. Shape parameters are defined to quantitatively represent ash shape by combining geometric values such as perimeter and area$^{8-14}$. Once the parameters are defined, the shape of an ash particle can be classified by calculating the parameters in an objective way that is free from human bias. Coupled with improvements in instruments for imaging ash particles and calculating shape parameters, quantitative analyses have become regarded as effective for classifying volcanic ash shape. For example, the Morphologi G3S$^{\mathrm{TM}}$ (Malvern Instruments$^{\mathrm{TM}}$) can rapidly measure the shape parameters and intensity of more than one thousand particles$^{15}$, which is very useful for volcanic ash analysis$^{9}$.

However, one problem with quantitative analyses by parameterization is the selection of the parameters used to represent the ash shape correctly$^{8}$, because the shape of volcanic ash is usually so complex that it cannot be described using only a single parameter. Although instruments such as the Morphologi can calculate parameters automatically, we still have to choose the shape parameters to be used. Thus far, many shape parameters, including fractal dimension, have been used for better understanding and classifying ash shapes$^{8-14}$. While quantitative analysis is objective once the parameters are decided, defining and selecting effective parameters is difficult. In order to assess the effective parameters, statistical analyses (multivariate and discriminant analyses) have been performed$^{8,13}$, which can reduce the number of parameters. However, prior to the statistical analyses, we have to select a limited number of shape parameters. In addition, it is uncertain whether the quality of the particles, such as complicated surface texture, is parameterized in sufficient detail.

Recently, in the field of machine learning, convolutional neural networks (CNNs) have been utilized for shape recognition with great success$^{16-18}$. A CNN contains layers that convolve the pixel intensities and send the convoluted signals to the next layer. Finally, the last (fully-connected) layer of the CNN outputs the probabilities that the object being evaluated falls into each class. Each layer contains weights that are updated during training by minimizing a loss function. If training is performed effectively, a CNN can often recognize objects with an accuracy comparable to that of humans. One advantage of CNNs is that we do not have to define complicated parameters to represent shape because the network automatically learns to categorize objects during training. Moreover, once the CNN is trained, we can share the network and use it to categorize new objects. Thus, CNNs have a large potential to improve volcanic ash analysis by eliminating the individual bias associated with visual analysis and the difficulty in choosing shape parameters that can effectively discriminate ash particle shapes. Here, we demonstrate this idea by classifying volcanic ash particles generated by different types of eruptions using a CNN.

\section*{Preparation of particle images}
\subsection*{Locations of samples}
In this work, we considered volcanic ash generated by three eruption types: magmatic, phreatomagmatic, and rootless (Table \ref{table_list}). Magmatic eruption is caused by the vaporization of volatiles contained in ascending magma by pressure release. If ascending magma makes contact with external water such as ground water, this type of eruption is called a phreatomagmatic eruption. While magmatic and phreatomagmatic eruptions are induced by magma ascending through the conduit, rootless eruption is caused by the interaction between lava and wet ground at the surface$^{19}$. When lava flows on the wet ground, water in the substrate vaporizes and the pressure rises dramatically between the lava flow and the substrate, which induces an explosion. Due to the repeated overlapping of active lava flow on water-saturated substrate, further explosions occur.

\begin{figure}[h]
\centering
\includegraphics[width=16cm, bb=0 0 2771 2172] {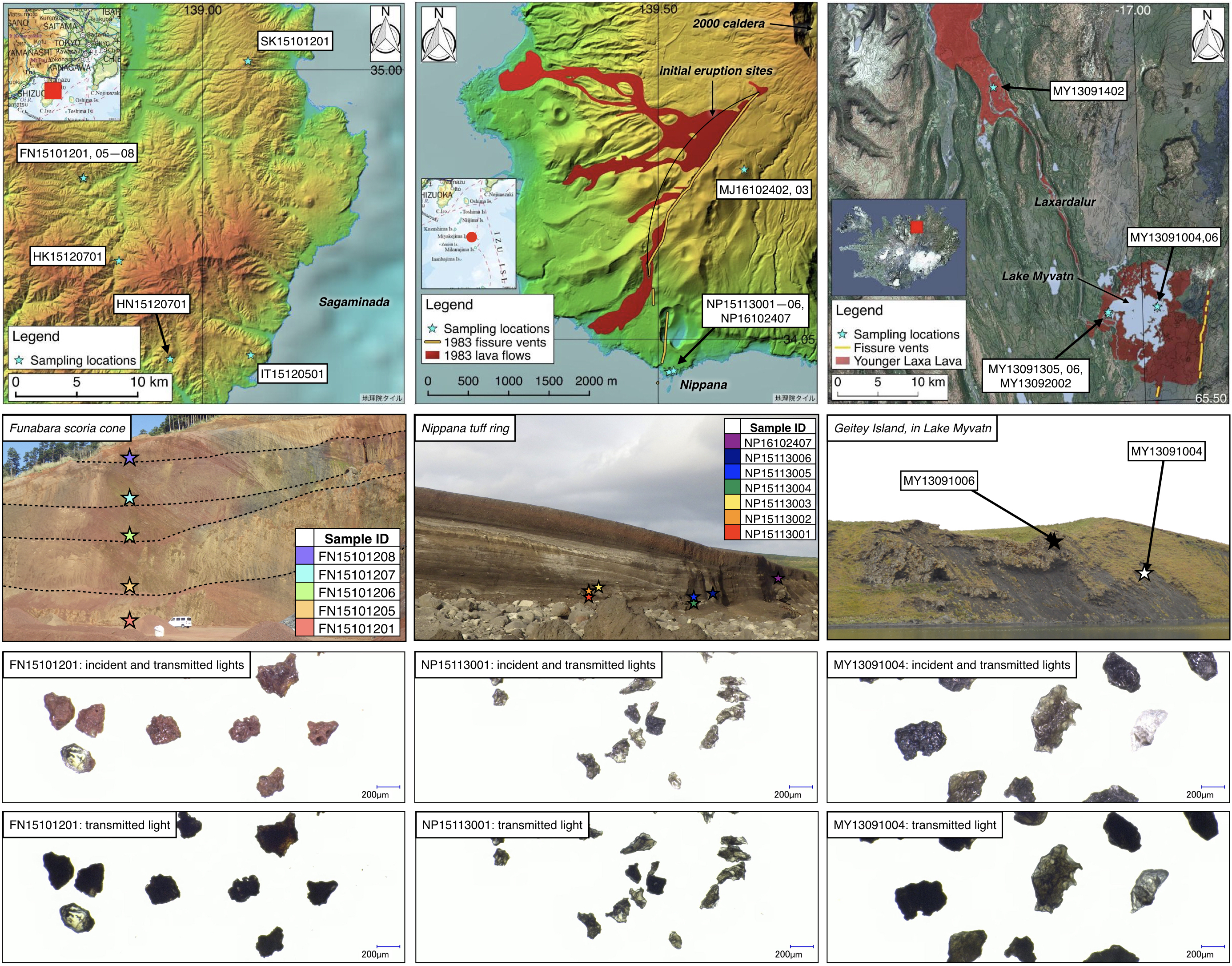}
\caption{Map, outcrop, and ash particle images for samples in this study (Left: Izu Paninsula, Japan, middle: Miyakejima Island, Japan, right: Myvatn, Iceland). Maps of Japan are based on the Chiriin Tile (http://maps.gsi.go.jp) of Geospatial Information Authority of Japan and the Hydrographic and Oceanographic Department, Japan Coast Guard. The 1983 lava extent is from "Miyakejima geological map 1:25,000" vector data of Geological Survey of Japan (https://gbank.gsj.jp/datastore/)$^{20}$. For the map of Iceland, the tiny black lines indicate the 10 m interval topographic contour lines based on the elevation model of the Landm\ae lingar \'Islands (LMI), and the background image is a Landsat image mosaic in RGB (resolution: 30 m), based on data from the National Land Survey of Iceland (NLSI).}
\label{map_outcrop_grain.jpg}
\end{figure}

The locations where the ash particles were collected are shown in Fig. \ref{map_outcrop_grain.jpg}. Every volcano in this study is a monogenetic basaltic volcano that has a similar bulk composition and phenocryst mode (Table \ref{Petrological}). Funabara, Hachikuboyama, Hachinoyama, Inatori, and Sukumoyama (Sample IDs begin with FN, HK, HN, IT and SK in Table \ref{table_list}, respectively) are scoria cones in Izu Peninsula, which were formed in the late Quaternary$^{23}$. 
Whole samples from Miyakejima (IDs begin with MJ and NP) were generated by the eruption in 1983. The eruption began from the flank of the mountain (Oyama) and then propagated to the summit and the coastal areas through a fissure vent system$^{33}$. Samples with IDs beginning with MJ and NP are fire fountain deposits and phreatomagmatic surge deposits (at an outcrop of Nippana tuff ring), respectively. Samples from Myvatn (IDs begin with MY) are products of explosive lava-water interactions i.e., rootless eruption, which occurred in 2170$\pm$38 cal yr BP$^{25}$. This explosive interaction created a conical edifice (known as a rootless cone) through the ejection of fragmented lava and water-saturated sediments. 

FN and NP samples were collected from several layers at the outcrops (Fig. \ref{map_outcrop_grain.jpg}). Except for MJ16102402 and MJ16102403, every particle was sampled from constituent materials of pyroclastic cones. Examples of microscope images of the volcanic ash particles are shown in Fig. \ref{map_outcrop_grain.jpg}. Each sample was sieved using screens after drying in an oven. Additional sample information, including particle size distributions and microscopic observations, is provided in the Supplementary Information.

\subsection*{Particle imaging}
Because a large number of particle images are required to train a CNN, we used the Morphologi G3S$^{\mathrm{TM}}$ (Malvern Instruments\texttrademark) to image volcanic ash particles$^{15}$. Morphologi is an automated particle analyzer that can detect the position and measure the size and shape parameters of thousands of particles in tens of minutes$^{15}$. The images taken by Morphologi are two-dimensional projections of particles dispersed on a glass plate. In this study, we used particles with grain sizes of 2$\phi$ to 3$\phi$ (125 $\mu$m to 250 $\mu$m in diameter) because they exist in sufficient quantities for training a CNN and statistical analysis. To place the volcanic ash particles on the glass plate, we used a sample dispersion unit (SDU) with an injection pressure of 1.5 bar and an injection time of 20 ms. We measured particles a 5$\times$magnification because this is the most suitable magnification to obtain the high-resolution shape of the particles$^{9}$. During the measurement, the illumination was set to diascopic (bottom lighting) under automatic light calibration (calibration intensity of 80.00 and intensity tolerance of 0.20). The threshold for background separation (0 to 255) was set at 80 in order to obtain a sharp focus. The measurement lasted approximately 40 min for each sample. Among the measured particle images, some are unsuitable for use and were excluded by visual inspection: images containing dust and images of overexposed, non-isolated, or cut-off particles. To train the CNN, we resized every image to 50$\times$50 pixels using Th-MakerX (http://www5.wind.ne.jp/miko/index-en.html).

\subsection*{Characteristic shape of particles}
\begin{figure}[h]
\centering
\includegraphics[width=20cm, bb=0 0 800 800] {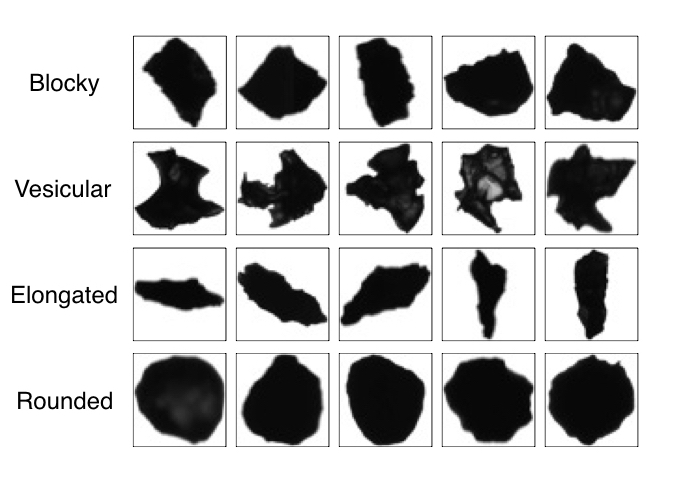}
\caption{Four basal shapes considered in this study, and examples of ash particles composed of a single basal shape.}
\label{base}
\end{figure}

In this study, since we aim to analyze the volcanic ash characteristics for each sample, it is necessary to classify ash into some category of shape. Previous studies have tried to classify volcanic ash particles into several categories: blocky$^{11,12,34}$, vesicular$^{8,12,34}$, mossy$^{11,34}$, shard$^{8}$, fluidal$^{11}$, and rounded$^{34,35}$. Considering these studies, we applied four typical basal shapes for volcanic ash particles: blocky, vesicular, elongated, and rounded (examples are shown in Fig. \ref{base}). Blocky particles are close to rectangular, with relatively straight and nearly right-angle edges. Vesicular particles have a concave and irregular shape. Elongated particles are relatively long and thin. Rounded particles have a relatively smooth and rounded shape caused by surface tension within a fluid droplet before they cool.

The origin of each particle shape depends on the eruption mechanism$^{8,12,36}$. Blocky and vesicular particles reflect fragmentation and quenching of magma, while elongated and rounded shapes are related to ductile deformation$^{35}$. In magmatic eruption, blocky particles are generally formed due to brittle fragmentation of magma in the volcanic conduit and on the ground surface. Vesicular particles are formed by rupturing of vesiculated magma in the conduit during explosive eruption. On the other hand, fragmented magma that has a low enough viscosity to deform forms either elongated particles due to stretching or rounded particles due to the effect of surface tension.
Phreatomagmatic and rootless explosions occur due to rapid heat transfer from magma to water, which produces explosive vaporization and results in brittle fragmentation$^{1,34}$. Blocky and vesicular particles are generated by the quenching of such magma fractures. On the other hand, ductile deformation of magma is caused by the fluid instabilities at the interface between magma and water$^{11,34}$. Elongated and rounded particles with a smooth surface are formed by the effect of surface tension during the time when the viscosity of magma is low. Rounded particles with a rough and cauliflower-like surface are formed when the surface of magma clasts cools faster than the effect of surface tension$^{37}$. Our samples consist of juvenile volcanic ash and phenocrysts, which did not receive weathering or hydrothermal alteration that could be counted as blocky or rounded. Most of the phenocrysts have a blocky shape (plagioclase, pyroxene, magnetite, and olivine).  

\section*{Training of CNN}
\subsection*{Mixing ratio determined by CNN}
Although the four classes defined above are the characteristic shapes of the particles in our samples, a single particle is often a mixture of more than two basal shapes. This is because volcanic ash particles are generated by multiple processes during fragmentation of magma and transport of pyroclasts, which has several components such as melts and crystals. Thus, there are many particles with a complex shape that should not be forced into a single class because not only does the classification become subjective but the properties of the neglected classes are discarded. 

One of the properties of a CNN is that it can output probabilities that a single particle falls into each class. For example, if a particle has primarily one basal shape, the corresponding class probability will be high and the probabilities of the other classes will be very low. On the other hand, in the case of particles composed of multiple basal shapes, two or more classes will have a relatively high probability. Thus, these probabilities can be used to quantitatively express the mixing ratio of basal shapes present in a single particle, which is an impossible task for humans. Therefore, we first trained the CNN using particles composed of only one of the four basal shapes being considered (i.e., pure shapes). Then, using the trained CNN, the probabilities for all four shapes were calculated for every particle in the samples.

\subsection*{Selection of basal images for training}
We visually selected images of particles that have only one basal shape, as shown in Fig. \ref{base}, without any qualitative thresholds. Of the approximately 15,000 ash images acquired, the numbers of images corresponding to particles composed of only blocky, vesicular, elongated and rounded shapes were 245, 239, 248 and 245, respectively (every particle image we selected is given at https://www.researchgate.net/publication/324495337\_Training\_particle
\_images\_for\_Classification\_of
\_volcanic\_ash\_particles\_using\_a\_convolutional
\_neural\_network
\_and\_probability). These basal images were used to train the CNN. 

Although the purpose of this work is to classify particle shape without having to select shape parameters, in order to verify the shape of the particles in the basal images, we measured the averaged characteristic shape parameters (aspect ratio, convexity, high sensitivity (HS) circularity, and solidity) of the basal images, as shown in Table \ref{parameter} (definition and method are shown in the Supplementary Information). Elongated particles have a low aspect ratio, and rounded ones have high circularity, which are consistent with the definitions of these classes. Vesicular particles are less circular compared with other classes because we selected particles that have large-scale concavities as the vesicular class. Particles of the blocky class have a larger aspect ratio and smaller circularity compared with those belonging to the elongated and rounded classes, respectively. Thus, the basal images we selected visually were distinguished relatively well.

Of these images, we used 200 images for each class as the training images, and the rest of the images were placed in the test set for evaluation of the network accuracy. The test images were not used for tuning of CNN. After the training of each epoch, the accuracy of the network was evaluated based on how many test images were categorized into the correct classes. Although we also calculated the accuracy using the training images, the accuracy by the test images is more important to evaluate the robustness of the CNN.

\subsection*{Structure and training of CNN}
\begin{figure}[h]
\centering
\includegraphics[width=18cm,bb=0 0 2470 1369] {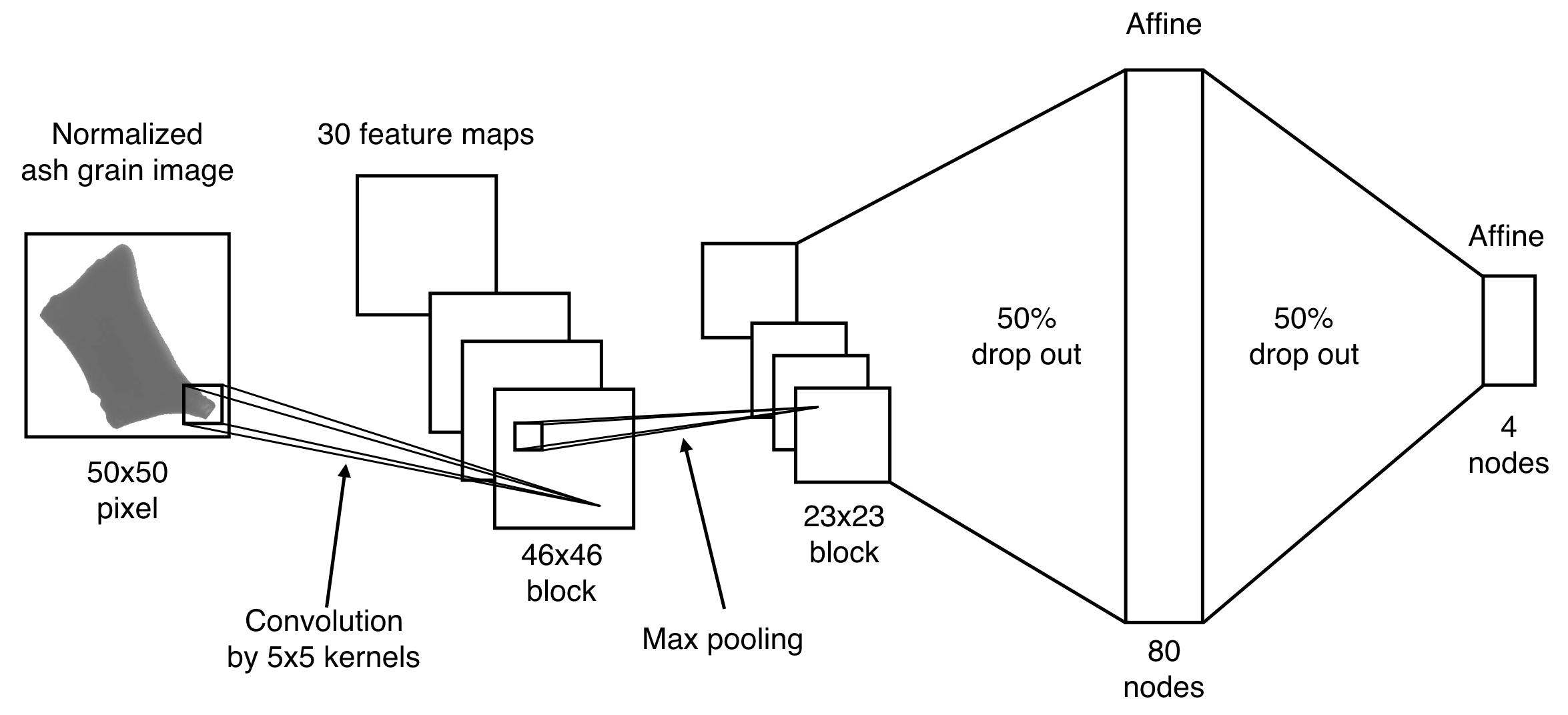}
\caption{Schematic view of CNN used for volcanic ash analyses.}
\label{structure}
\end{figure}

Training and testing were performed using the Keras package (https://keras.io), which is free and written in Python$^{38}$. The structure of the CNN used in this study is shown in Fig. \ref{structure}. The images of the volcanic ash particles used were 50$\times$50 pixels, and the pixel intensities were normalized to between 0 and 1 and used as the input signal of the CNN. First, the pixel intensity values are convoluted using 5$\times$5 kernels. The number of kernels was set to 30, and thus 30 convolution maps composed of 46$\times$46 blocks are generated. Although we tried training several models, and specifically investigated the effect of image padding (including extra space around the images), the accuracy changed little because our network has only one convolutional layer and the detailed features around the image boundary are not so important for the four basal shapes. Thus, image padding, which is important for deeper networks with many convolutional layers, was not considered in this work. The convoluted maps are pooled using a 2$\times$2 max pooling layer, which results in 23$\times$23 block maps. The pooling layer, which removes the effect of trivial movement of objects in images and increases the robustness of the network to displacements of objects, has a stride equal to the length of the filter and selects the maximum value in each area (2$\times$2 block). After the pooling, the signals are sent to an affine layer, and then the probabilities for each of the four classes are calculated. 

One of the most serious problems for neural networks is overfitting, which happens when a network is tuned only for the training data and does not generalize well to unseen data (accuracy is high only on the training set). In the present study, the number of training images is small, which is a common cause of overfitting. Thus, in order to avoid overfitting, we added 50\% dropout layers before the affine layers. A dropout layer randomly severs a specified percentage of interlayer connections of the network during each training iteration, which reduces the complexity of the network and thus prevents overfitting$^{39}$. In the testing phase, although every connection is used, signals are scaled by the ratio of the cutting of the connection$^{39}$.
The mini-batch size was set to 200 (in one epoch, 200 images are selected randomly from the 800 training images) and the number of iterations in one epoch to update the weights of each layer was 20 (updating of weights using 200 mini-batch images was conducted 20 times per epoch)$^{38}$. The initial values of the weights were set using a Glorot uniform initializer, in which numbers are distributed uniformly within the range depending on the layer number$^{40}$. Gradient descent with an Adam optimizer$^{41}$ was used to update the weights. The probabilities and loss function were calculated using the softmax and cross-entropy functions, respectively, in the Keras library.

Because the number of images was small, in order to avoid overfitting, 200 images in one mini-batch selected in one epoch were rotated, flipped horizontally and shifted in one of four directions randomly at the start of each epoch. This procedure can increase the number of the training images virtually, even though the exact number does not change from 800. The range of the rotation angle was -180 degrees to 180 degrees, and the maximum amount of shift was 10\%. One image was augmented by the combination of these operations in Keras$^{38}$. The angle and width of rotation and shift were determined randomly within the maximum range, and flipping occurred randomly. Thus, one epoch corresponds to selection of one mini-batch, data augmentation, and then 20 iterations of weight updates. After one epoch, a new mini-batch was selected and the same process was conducted. We trained the CNN until the loss function and the accuracy became stable.

\section*{Results}
\subsection*{Accuracy of CNN}
\begin{figure}[h]
\centering
\includegraphics[width=18cm,bb=0 0 2470 1369] {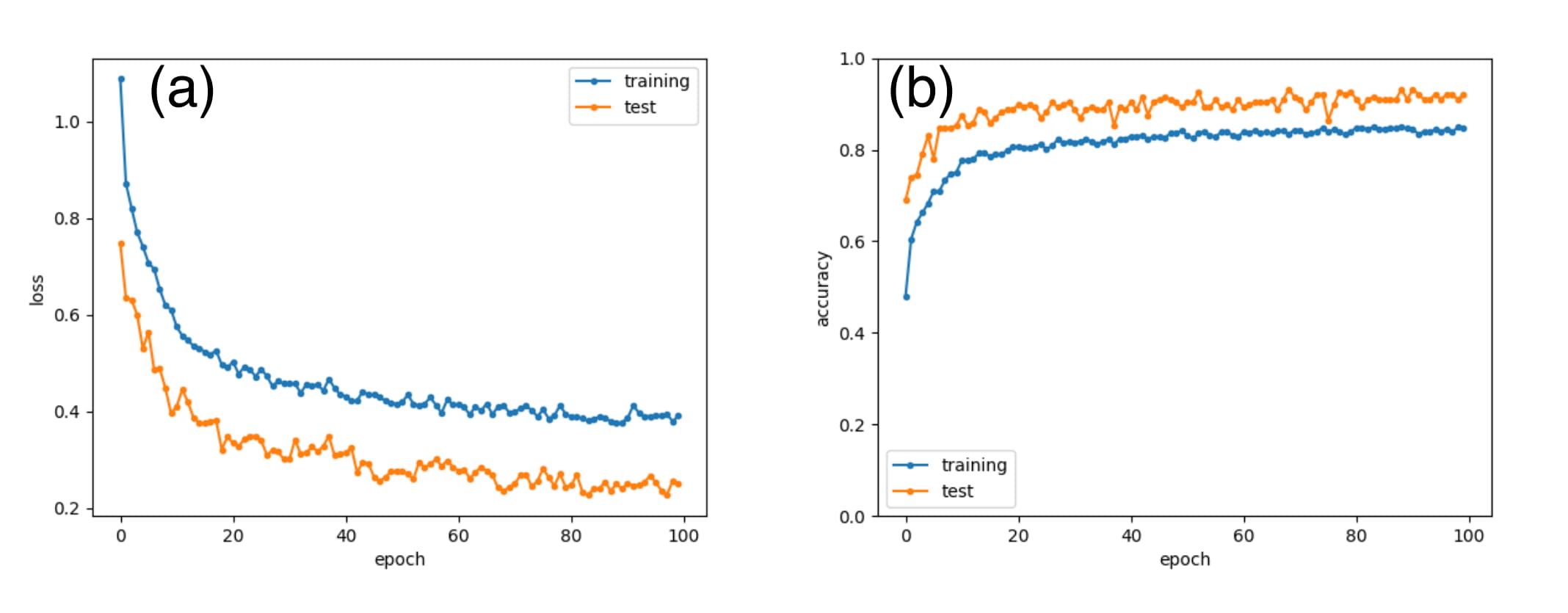}
\caption{(a): Loss functions and (b): training and test accuracies as a function of epoch number.}
\label{loss}
\end{figure}

The loss function and network accuracy as a function of epoch are shown in Fig. \ref{loss}. As the epoch number increases, the loss functions using mini-batch decrease and the training and test accuracies increase. From the 80th epoch, the loss function becomes stable, so we decided to run 100 epochs. During training, the training loss is higher than the testing loss (Fig. \ref{loss}). This is because the regularization mechanism of the dropout layer is turned off at testing time$^{38}$. We confirmed that the training loss and the testing loss (and the accuracies) became the same if the two dropout layers are removed. The accuracy of the CNN after 100 epochs was evaluated from test images to be approximately 92\% (Fig. \ref{loss} b). 
Because initial values of layers and the mini-batch were determined randomly, we trained the CNN several times, and confirmed that the corresponding changes in accuracy were negligible. 

We also conducted training with a small number of the training images (the number of the test images did not change). In the case of 100 images per class (400 images total), the accuracy of the network was approximately 90\%. Approximately 90\% accuracy could often be achieved even with 50 images per class, and it decreased to $\sim$86\% at 20 training images per class. Thus, the minimum number required for classification of our data set is 50 per class (200 total) in order to achieve $\sim$90\% accuracy. Because the images taken by Morphologi have a relatively simple shape (only four classes and the complicated surface texture was not shown), a sufficient number of training images were generated by augmentation (rotating, shifting, and flipping) even though the number of the original images was small. When the training data was not augmented, the accuracy was only $\sim$80\%, even with 200 images per class, and overfitting occurred. Thus, in the case of data without augmentation, many more images (perhaps more than one thousand per one class) are needed. We note that the required number of images strongly depends on the structure of the network and the images themselves. If we were to consider the complicated patterns in, for example, scanning electron microscopy (SEM) images, many images would be required, which is beyond the scope of this work.

\subsection*{Mixing ratio of particles}
\begin{figure}[h]
\centering
\includegraphics[width=20cm, bb=0 0 1000 1000] {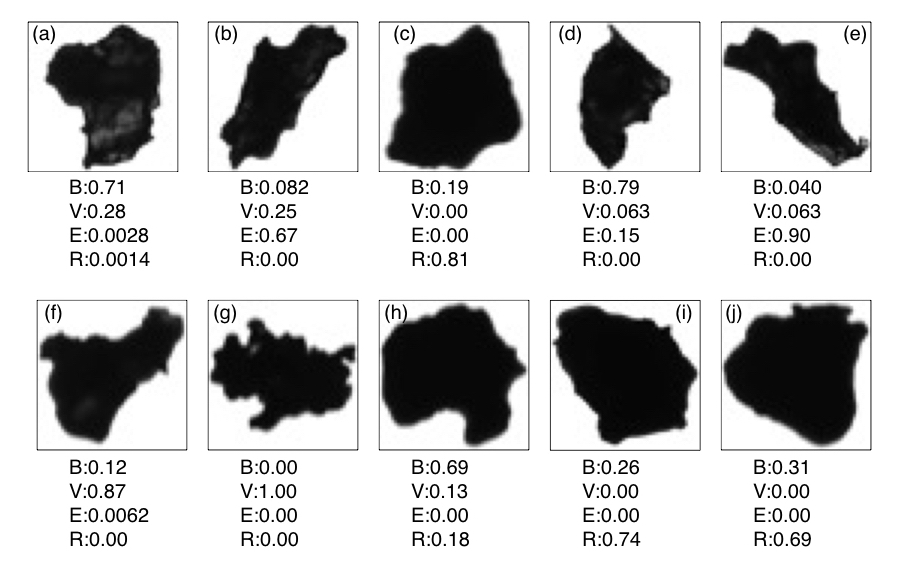}
\caption{Examples of ash particles with complex shapes and corresponding class probabilities for each of the four basal shapes: B (blocky), V (vesicular), E (elongated) and R (rounded). The probabilities can be interpreted as the mixing ratio of the four basal shapes.}
\label{ratio}
\end{figure}

Using the trained network, we calculated the probabilities for each of the four classes for every particle in the data set. Examples of probabilities calculated by the trained CNN are shown in Fig. \ref{ratio}. If these particles were primarily composed of two basal shapes, the probabilities for two classes were relatively high, while only one probability is high for the particles composed of only one basal shape (e.g., Fig. \ref{ratio} g). In the case of Fig. \ref{ratio} h, three shapes are mixed, so the probability of the vesicular class was relatively lower compared with the particles in Fig. \ref{ratio} a and b. 
However, we note that not every combination of classes was observed in our analyses. The rounded class was found to mainly mix with the blocky class. In addition, the number of particles composed of three basal shapes (e.g., Fig. \ref{ratio} h) was small. The rounded particles selected as training images have a high aspect ratio and circularity, which is in contrast to the elongated and vesicular shapes (Table \ref{parameter}), i.e., the rounded shape is exclusive from vesicular and elongated shape by definition of the classes. Thus, the probability of the rounded shape was not often found in combination with the vesicular and elongated shapes with high probability.

In this work, the training images were selected from the large-scale structure of the particle, and thus some basal images contain small-scale structure such as concavity. Thus, uncertainty due to the effect of the small-scale structure could not be removed completely in evaluating the mixing ratio. For example, even though some particles have shallow or small concavity (e.g., Fig. \ref{ratio} e, d and j), the vesicular class probability for these particles was low. This is because we defined the vesicular shape to be an irregular outline and concavity that can be recognized as a whole structure rather than part of it  (Fig. \ref{base}). Thus, the probabilities by our CNN show the mixing ratio of large-scale structure in one particle. If we limit the training images to more ideal shapes (only particles without small-scale structures), accurate evaluation of small vesicular shape may be possible. However, because volcanic ash is not an artificial object, many more samples are required to collect better training images. Moreover, in the case of Fig. \ref{ratio} d, the CNN might recognize that the small concavity is the connection between blocky and elongated shapes rather than vesicular structure, which also reduced the vesicular probability. Another caveat is that we used relatively low-resolution (50$\times$50 pixel) images to reduce the computational time. Thus, the CNN might not recognize the small-scale structure of the particles. If high-resolution images are used, although the computational time will increase, the effect of detailed morphology may be evaluated more accurately. These tasks require more images and can be considered in future studies.

\subsection*{Averaged mixing ratio and clustering of samples}
As a next step, we selected 18 samples from three areas, Funabara, Nippana, and Myvatn (formed by magmatic, phreatomagmatic, and rootless eruptions, respectively), and performed clustering of the samples based on the mean probabilities of the four basal shapes averaged over every particle in each sample. Moreover, the probability ratio of basal shapes over the three areas $p_i$ was also calculated as $\displaystyle p_i=\sum_{j} m_{ij}n_j/\sum_j n_j$, where $m_{ij}$ and $n_j$ are the mean probability of basal shape $i$ (one of four classes) in sample $j$ and the number of particles in the sample, respectively. The summation was carried out for each area (Funabara, Nippana, and Myvatn).

One caveat is that, with Morphologi images, details of the surface texture cannot be observed. However, Morphologi can automatically calculate the pixel intensity for each particle$^{15}$, which can provide rough information about the particle composition in terms of the transparency under bottom lighting.
Transparency can be an effective means of extracting glass fragments and transparent crystals from volcanic ash, which can play an important role in a discussion of the degree of quenching and fragmentation of magma/lava$^{3,42,43}$. To calculate ash transparency, complex parameterization is not required. Thus, we considered the transparency in addition to the class probabilities. 

Using Morphologi, we evaluated the mean ($I_m$) and standard deviation ($I_{SD}$) of the intensity of one particle (see the Supplementary Information for details), and thus there are 2 parameters per particle image. We averaged $I_m$ and $I_{SD}$ over one sample, and considered the averaged value, the standard deviation, and the median as the transparency of each sample. Thus, each sample has 4 parameters (averaged probabilities) for shape and 6 parameters for transparency. Cluster analyses were conducted using these 10 parameters. We performed hierarchical clustering (Ward's method$^{44}$) in R$^{45}$ based on the Euclidean distances between each pair of samples.

\begin{figure}[h]
\centering
\includegraphics[width=10cm, bb=0 0 1000 1000] {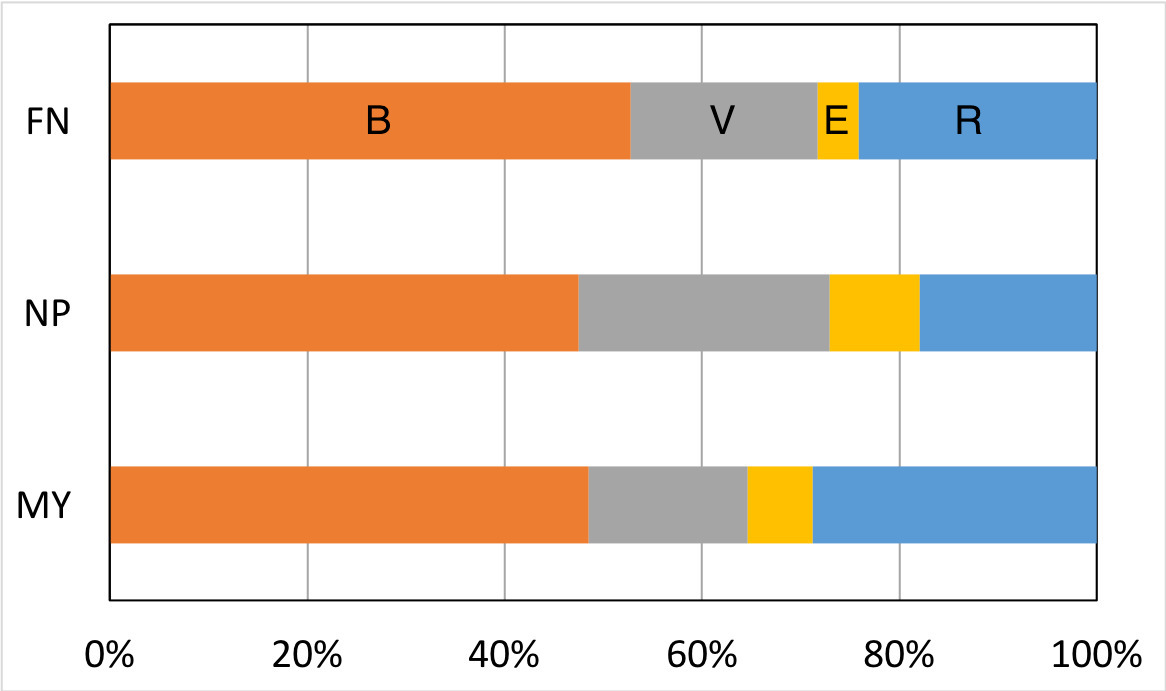}
\caption{Mean shape probabilities over three areas (three eruption types). B: blocky, V: vesicular, E: elongated and R: rounded.}
\label{bar}
\end{figure}

Table \ref{table_data} shows the mean probabilities of basal shapes and transparency values for each sample. The averaged probability over each area (three different eruption types) $p_i$ is shown in Fig. \ref{bar}. In every sample, the most dominant shape was blocky. In particular, particles of Funabara (FN) consist of more than 50\% blocky shape on average (Fig. \ref{bar}). The probabilities of the vesicular and elongated shapes were high (more than 30\% and 10\%, respectively) for the lower layers of Nippana (NP15113001--NP15113003). The MY13091305 particles also had a high probability for the elongated shape. On the other hand, the probability for the rounded shape is small for the Nippana samples (Fig. \ref{bar}). In addition to vesicularity and elongation of particles, transparency is also high for the Nippana samples (there are many transparent particles in the Nippana samples) compared with the samples of Funabara (FN) and Myvatn (MY) (Table \ref{table_data}) because most of the particles in the Nippana samples are glass fragments (see the Supplementary Information). 
 
The reason why the blocky shape is dominant is that blocky particles are common in shape. As shown above, blocky particles are general products in brittle fragmentation of magma, which is typical for magmatic eruption. The probability of more than 50\% for the blocky shape for the Funabara samples can be explained in this way. Furthermore, the dominance of the blocky shape in the Nippana and Myvatn samples can be explained as follows: blocky, equant shapes are the most frequently found in hydrovolcanic (magma-water interactions) ash$^{34}$. The high probability of the vesicular shape in the lower layers of Nippana (NP15113001--NP15113003) is consistent with general understanding for basaltic phreatomagmatic products: fragmentation due to thermal shock and expansion of gas by the contact of vesiculating basaltic magma and external water produces vesicular glassy ash particles$^{34}$. The high ratio for the elongated shape in the lower layers of Nippana (NP15113001--NP15113003) and MY13091305 indicates the existence of fluidal shape particles (e.g., Supplementary Fig. S6), which is caused by the elongation of low-viscosity magma during eruption. The effect of elongated crystals can be excluded because the samples in this work were formed in a short duration, in which case it is difficult to vary the phenocryst component and proportions. The dominance of rounded particles in the Myvatn samples may correspond to the existence of "mossy grain", which was described for Icelandic rootless tephra$^{11}$. Mossy morphology is formed by hydrodynamic fragmentation of deformable lava if the effects of viscosity are more dominant than the effects of surface tension$^{11}$. In the case of MY13091402, the high ratio of the rounded shape is consistent with the existence of hollow spherules (see Supplementary Fig. S8 and the Supplementary Information for details). For FN15101206, FN15101208, and NP16102407, the reason for the high rounded shape probability may be surface tension $^{11,37}$.

\begin{figure}[h]
\centering
\includegraphics[width=16cm, bb=0 0 1000 1000]{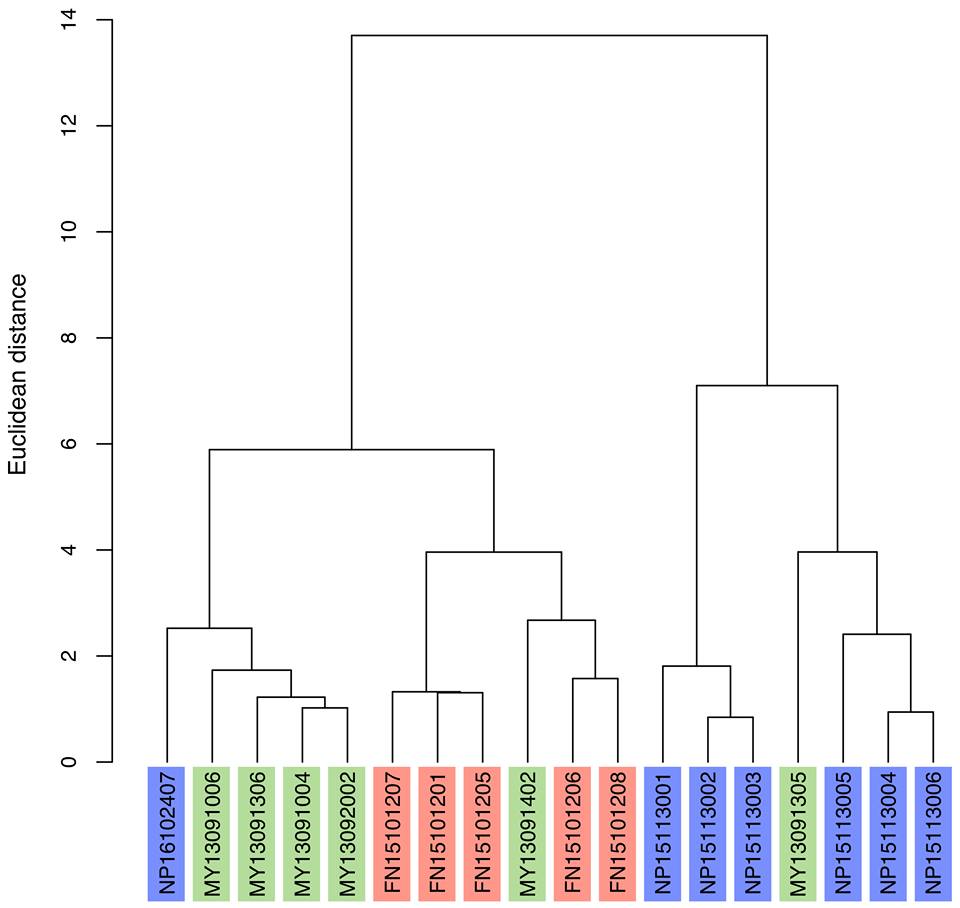}
\caption{Dendrogram of samples using averaged probabilities and transparencies. Sample IDs shown with red, blue, and green indicate Funabara (magmatic eruption origin), Nippana (phreatomagmatic eruption origin) and Myvatn (rootless eruption origin), respectively. }
\label{cluster}
\end{figure}

The clustering of samples is shown in Fig. \ref{cluster}. The clusters are consistent with the differences between the three regions (differences in eruption type). In the dendrogram, first, most Nippana samples were distinguished from the others, then the Funabara and Myvatn samples were separated. The clusters are also consistent with the stratigraphy: NP15113001--NP15113003 (lower parts of Nippana tuff ring), and NP15113004--NP15113006 (upper parts of Nippana tuff ring). This is also the case for the Funabara samples, although FN15113007 does not follow.

NP16102407, MY13091305, and MY13091402 were not classified consistently with their eruption type. However, this can be explained by the differences in the stratigraphic positions and geospatial distributions of the sampling. MY13091305 was collected from the lower part of a rootless cone (Fig. \ref{map_outcrop_grain.jpg}), and contains information about the early stage of lava-water interaction, which is considered to possibly be as energetic as a phreatomagmatic explosion$^{46}$. MY13091402 was collected in Hagi, where the lava flowed down approximately 45 km from the fissure vent (Fig. \ref{map_outcrop_grain.jpg}). Therefore, this sample may be different from other samples of rootless eruptions due to different conditions, such as lava temperature and availability of water during the explosions. NP16102407 was separated from other phreatomagmatic samples (Fig. \ref{map_outcrop_grain.jpg}). It is known that, at the time of the formation, a magmatic eruption took place concurrently at an extremely close range (less than 300 m)$^{33}$. Thus, NP16102407 might be a mixture of ash particles generated by phreatomagmatic and magmatic eruptions, and is therefore considered to be distinct from other phreatomagmatic samples.

\section*{Discussion}
\subsection*{Future works}
In this work, because Morphologi was used to quickly acquire many images, the detailed surface texture of the particles, other than the transparency, could not be imaged. Surface texture also contains information about eruption type, as both qualitative and quantitative analyses have indicated$^{1,4,5,47}$. Thus, we are planning to apply this method to other types of images, such as SEM images, which provide such detail$^{3,5,47}$, with the goal of training our CNN to recognize particle texture. 

In order to increase the robustness of the particle classification by CNN, much more images should be prepared and tested with several conditions (e.g., different location, resolution and class). As we mentioned, our basal images were selected from large structure. More accurate mixing ratios may be obtained if the training images are limited to more ideal shapes, which do not have small-scale structure. However, the number of particles with ideal shapes is very small. To solve the problem of the amount of images, a database of volcanic ash images$^{9}$ will be useful for further studies. The structure of CNN also should be discussed especially when the complicated texture of particles is taken into account. 

\subsection*{Suggestion of new taxonomy approach}
Other than volcanic ash, our approach using probabilities (mixing ratio) is applicable to classifications of other objects in nature, such as organisms and fossils that are difficult to categorize into unique species$^{48}$. Although an important principle of classification is that one object is categorized into one class and that every class is independent, in the case of natural objects, characteristics of several classes are often present in a single object. Thus, classification by the naked eye becomes subjective if we forcibly assign such natural objects into a single class. On the other hand, visual classification has the advantages that qualitative characteristics can be taken into account and that unlimited parameterization of complex shapes is not required. As quantitative analyses of volcanic ash have shown$^{8}$, the number of parameters increases as one seeks a better representation of complex objects. A CNN can provide probabilities for each class (i.e., quantification of human recognition), which can be interpreted as a mixing ratio of basal shapes. Thus, in addition to the fact that a trained CNN can be shared, even obscure objects can be classified without forcibly assigning them into a single class and without defining numerous complex parameters, which may become a new direction of taxonomy. Further studies and discussions with researchers in different fields (e.g., geology, computer science, and biology) are important.

\section*{Acknowledgements}
The sample collections were supported by the Izu Peninsula Geopark Promotion Council and the Sasakawa Scientific Research Grant from the Japan Science Society for R.N(25-602). D.S. is supported by a JSPS research fellowship. H.H. is supported by JSP CREST JPMJCR1761. We used Morphologi G3S at the Geological Survey of Japan, AIST. R.N. and H.H. are supported by KAKENHI No. 17H02063 and the Joint Usage/Research Center Program No. 2015-B-04 from the Earthquake Research Institute, University of Tokyo.

\section*{Author contributions}
D.S. performed the training and the testing of the CNN. R.N. collected volcanic ash and took pictures of each particle. D.S., R.N. and S.O. made data sets of ash images discussing the shapes of particles. The structure of the network was discussed by D.S. and H.H. All authors prepared the manuscript. 
\section*{Additional information}
The authors declare no competing interests outlined in the editorial policies.


*Title etc. translated by R.N.

\newpage

\begin{table}[htbp]
\begin{center}
\caption{List of samples. }
\scalebox{0.7}{
\begin{tabular}{|l|l|l|l|l|l|l|l|l|l|}
 \hline
Eruption type & Sampling location & Sample ID & Number of ash particles \\
\hline
Magmatic & Funabara, Izu Peninsula & FN15101201 & 131\\
& Funabara, Izu Peninsula & FN15101205 & 262 \\
& Funabara, Izu Peninsula & FN15101206 & 206 \\
& Funabara, Izu Peninsula & FN15101207 & 87 \\
& Funabara, Izu Peninsula & FN15101208 & 168 \\
& Hachikuboyama, Izu Peninsula & HK15120701 & 196  \\
& Hachinoyama, Izu Peninsula & HN15120701 & 143 \\
& Inatori, Izu Peninsula & IT15120501 & 230 \\
& Sukumoyama, Izu Peninsula & SK15101201 & 230\\
& Miyakejima & MJ16102402 & 284\\
& Miyakejima & MJ16102403 & 421 \\
\hline
Phreatomagmatic & Nippana, Miyakejima & NP15113001 & 1851 \\
& Nippana, Miyakejima & NP15113002 & 707 \\
& Nippana, Miyakejima & NP15113003 & 428 \\
& Nippana, Miyakejima & NP15113004 & 1125 \\
& Nippana, Miyakejima & NP15113005 & 708 \\
& Nippana, Miyakejima & NP15113006 & 796 \\
& Nippana, Miyakejima & NP16102407 & 863 \\
\hline 
Rootless & Myvatn, N Iceland & MY13091004 & 923 \\
& Lake Myvatn, N Iceland & MY13091006 & 1065 \\
& Lake Myvatn, N Iceland & MY13091305 & 686 \\
& Lake Myvatn, N Iceland & MY13091306 & 670 \\
& Hagi, Myvatn, N Iceland & MY13091402 & 1479 \\
& Lake Myvatn, N Iceland & MY13092002 & 965 \\
 \hline
\end{tabular}
}
\label{table_list}
\end{center}
\end{table}

\newpage
\newpage

\begin{landscape}
\begin{table}[htbp]
\begin{center}
\caption{Petrological information of each volcano.}
\scalebox{0.7}{
\begin{tabular}{|p{3.7cm}p{1.5cm}|p{3cm}|p{3cm}|p{3cm}|p{3cm}|p{3cm}|p{3cm}|p{3cm}|}
\hline 
Name & & Funabara & Hachinoyama & Hachikuboyama & Inatori & Sukumoyama & Miyakejima 1983 & Myvatn\\
\hline 
Formation age & & 200$\pm$8 ka$^{21}$ & 36 ka$^{22}$ & 17 ka$^{23}$ & 19 ka$^{22}$ & 131 ka$^{22}$ & 1983 A.D.$^{24}$ & 2170$\pm$38 cal yr BP$^{25}$\\
\hline 
Bulk composition & SiO$_2$ & 50.91 & 50.24 & 51.61 & 49.32 & 49.02 & 52.91 & 49.3\\
(whole rock) &TiO$_2$ & 1.08 & 1.07 & 0.92 & 0.92 & 0.89 & 1.42 & 1.1\\
&Al$_2$O$_3$ & 17.44 & 19.64 & 18.87 & 17.62 & 18.1 & 14.98 & 14.98\\
&Fe$_2$O$_3$ & 3.61 & 2.98 & 1.31 & 3.34 & 1.68 & 14.33 & 0.9\\
&FeO & 6.51 & 5.82 & 6.63 & 6.25 & 8.25 & N/A & 9.27\\
&MnO & 0.17 & 0.15 & 0.16 & 0.17 & 0.16 & 0.24 & 0.19\\
&MgO & 6.88 & 6.33 & 6.71 & 7 & 7.55 & 4.13 & 6.96\\
&CaO & 9.63 & 9.81 & 9.42 & 12.22 & 10.48 & 9.11 & 12.63\\
&Na$_2$O & 2.62 & 3.21 & 3.03 & 2.36 & 2.43 & 2.67 & 2.15\\
&K$_2$O & 0.39 & 0.26 & 0.46 & 0.27 & 0.35 & 0.53 & 0.91\\
&H$_2$O$+$ & 0.29 & 0 & 0.02 & 0 & 0.25 & N/A & 0.79\\
&H$_2$O$-$ & 0.69 & 0.32 & 0.34 & 0.54 & 0.71 & N/A & 0.00\\
&P$_2$O$_5$ & 0.24 & 0.03 & 0.21 & 0.14 & 0.1 & 0.14 & 0.28\\
&Total & 100.46 & 99.86 & 99.69 & 100.15 & 99.97 & 100.46 & 99.46\\ 
\small{Reference}& & 26& 26 & 26 & 26 & 26 & 24 & 27\\
\hline 
Phenocrysts & & Ol, Pl, Px (A and hypersthene), Qz & Ol, Pl (clear and dusty), Cpx, Qz & Ol, Pl, Cpx & Ol, Pl (clear and dusty), Cpx & Ol, Pl & Pl, Cpx, Mt & Ol, Pl, Px\\ 
\small{Reference}& & 28 & 29 & 30 & 29 & 29 & 24 & 31\\
\hline 
Mode of phenocrysts &(\%) & 5.8\% (5.8\% of Ol, 0.2\% of Cpx)\% & 11.9--15.7\% (1.8\% of Ol, 5.7\% of clear Pl, 5.8\% of dusty Pl, and 0.9\% of Cpx) & 19.7\% (3.3\% of Ol, 6.3\% of Pl, 0.4\% of Cpx, 0.3\% of Qz)& 5.5--7.3\% (0.8\% of Ol, 2.7\% of clear Pl, 1.2\% of dusty Pl, and 1.2\% of Cpx) & 3.0--5.9\% (2.5\% of Ol, and 1.2\% of Pl) & 5 to 8\% of Pl, $<$ 0.1 to 0.3\% of Cpx, and $<$ 0.2\% of Mt & less than 10\%\\ 
\small{Reference}& & 30 & 29 & 30 & 29 & 29 & 24 & 30\\
\hline 
Phenocrysts size & & Ol: 1 mm (mode), Pl: 1 mm (mode), hypersthene: $<$ 1 mm, Qz: 3 mm (occasionally, in maximum) & In maximum: Ol: 2 mm, clear Pl: 2 mm, dusty Pl: 4 mm, and Qz: 1 mm & N/A & In maximum: Ol: 2 mm, clear Pl: 2 mm, dusty Pl: 2.5 mm, Cpx: 0.8 mm, and Opx: 0.5 mm & In maximum: Ol: 1 mm, Pl: 0.5 mm & Pl: 0.6 to 0.8 mm (major axes), 0.2 to 0.3 mm (minor axes), Cpx: 0.2 to 0.4 mm, Mt: 0.2 mm & Pl: 2 mm to 6 mm (1.5 to 2 cm for large)\\ 
\small{Reference}& & 28 & 29 & - & 29 & 29 & 24 & 32\\
\hline 
\end{tabular}
}

\small{Ol=olivine, Pl=plagioclase, Cpx=clinopyroxene, Mt=magnetite, A=augite, Qz=quartz}

\label{Petrological}
\end{center}
\end{table}
\end{landscape}

\newpage

\begin{table}[htbp]
\begin{center}
\caption{Averaged shape parameters (aspect ratio, convexity, high sensitivity (HS) circularity, and solidity) for each basal shape selected for training and testing of CNN. Errors represent standard deviation.}
\begin{tabular}{|l|l|l|l|l|}
\hline
 & Aspect ratio & Convexity & HS circularity & Solidity\\
 \hline
Blocky & 0.71$\pm$0.13 & 0.99$\pm$0.02 & 0.71$\pm$0.06 & 0.85$\pm$0.02\\
\hline
Vesicular & 0.69$\pm$0.13 & 0.87$\pm$0.05 & 0.48$\pm$0.08 & 0.73$\pm$0.09\\
\hline
Elongated & 0.37$\pm$0.06 & 0.96$\pm$0.09 & 0.51$\pm$0.08 & 0.80$\pm$0.23\\
\hline
Rounded & 0.80$\pm$0.09 & 0.99$\pm$0.01 & 0.82$\pm$0.04 & 0.89$\pm$0.01\\
\hline
\end{tabular}
\label{parameter}
\end{center}
\end{table}

\newpage

\begin{table}[htbp]
\begin{center}
\caption{Mean probabilities for four basal shapes, and average, standard deviation (SD) and median of I$_{\textrm{m}}$ and I$_{\textrm{SD}}$ over each sample.}
\scalebox{0.7}{
\begin{tabular}{|c|cccc|cccccc|}
\hline
Sample ID & Blocky & Vesicular & Elongated & Rounded & Ave. I$_{\textrm{m}}$ & SD I$_{\textrm{m}}$  & Median I$_{\textrm{m}}$  & Ave. I$_{\textrm{SD}}$  & SD I$_{\textrm{SD}}$  & Median of I$_{\textrm{SD}}$ \\ \hline
FN15101201  & 0.528 & 0.265 & 0.0458 & 0.161 & 15.160  & 2.067  & 14.972 & 12.939  & 1.361  & 12.882 \\
FN15101205  & 0.538 & 0.206 & 0.0570 & 0.199 & 15.203  & 1.882  & 15.175 & 12.377  & 1.257  & 12.319 \\
FN15101206  & 0.505 & 0.164 & 0.0279 & 0.303 & 15.149  & 1.770  & 14.986 & 12.532  & 1.395  & 12.500 \\
FN15101207  & 0.499 & 0.253 & 0.0636 & 0.184 & 13.928  & 1.763  & 13.941 & 12.528  & 1.367  & 12.326 \\
FN15101208  & 0.556 & 0.100 & 0.0228 & 0.322 & 15.137  & 1.780  & 15.164 & 12.765  & 1.384  & 12.682 \\ 
NP15113001 & 0.417 & 0.362 & 0.103 & 0.118 & 23.867  & 5.629  & 22.900 & 15.614  & 3.681  & 14.773 \\
NP15113002  & 0.434 & 0.332 & 0.129 & 0.106 & 23.043  & 5.020  & 22.268 & 15.122  & 3.423  & 14.246 \\
NP15113003  & 0.442 & 0.343 & 0.141 & 0.0737 & 22.114  & 4.628  & 21.587 & 15.071  & 3.024  & 14.368 \\
NP15113004  & 0.523 & 0.150 & 0.0803 & 0.246 & 21.387  & 4.876  & 20.379 & 13.448  & 2.547  & 12.875 \\
NP15113005  & 0.487 & 0.272 & 0.0826 & 0.158 & 20.874  & 5.007  & 19.855 & 14.084  & 3.209  & 13.308 \\
NP15113006  & 0.538 & 0.180 & 0.0800 & 0.202 & 21.131  & 5.057  & 19.967 & 13.735  & 2.879  & 13.086 \\
NP16102407  & 0.516 & 0.107 & 0.0420 & 0.334 & 19.815  & 5.246  & 18.810 & 12.738  & 2.607  & 12.287 \\ 
MY13091004  & 0.509 & 0.148 & 0.0721 & 0.271 & 16.636  & 6.110  & 15.148 & 12.545  & 2.902  & 12.118 \\
MY13091006  & 0.497 & 0.175 & 0.0645 & 0.264 & 15.166  & 3.926  & 14.484 & 12.568  & 1.959  & 12.409 \\
MY13091305  & 0.466 & 0.178 & 0.168 & 0.189 & 20.335  & 7.622  & 18.003 & 14.051  & 3.812  & 13.224 \\
MY13091306  & 0.481 & 0.146 & 0.0661 & 0.307 & 16.015  & 4.931  & 15.014 & 12.385  & 2.715  & 12.075 \\
MY13091402  & 0.448 & 0.177 & 0.0237 & 0.351 & 14.881  & 2.739  & 14.499 & 12.827  & 1.577  & 12.748 \\
MY13092002  & 0.516 & 0.135 & 0.0548 & 0.285 & 16.577  & 5.055  & 15.291 & 12.867  & 2.639  & 12.489 \\ \hline
\end{tabular}
}
\label{table_data}
\end{center}
\end{table}

\newpage

\begin{thebibliography}{}
\item
Heiken, G., \& Wohletz, K. Volcanic ash. Harvard \& MIT: University Presses of California, (1985).
\item
Gonnermann, H. Magma fragmentation. \textit{Annu. Rev. Earth Planet. Sci.} \textbf{43}, 431--458 (2015).
\item
Taddeucci, J., Pompilio, M. \& Scarlato, P. Conduit processes during the July--August 2001 explosive activity of Mt. Etna (Italy): inferences from glass chemistry and crystal size distribution of ash particles. \textit{J. Volcanol. Geotherm. Res.} \textbf{137}, 33--54 (2004).
\item
Ersoy, O., Chinga, G., Aydar, E., Gourgaud, A., Cubukcu, H. E. \& Ulusoy, I. Texture discrimination of volcanic ashes from different fragmentation mechanisms: A case study, Mount Nemrut stratovolcano, eastern Turkey. \textit{Computer \& Geosciences} \textbf{32}, 936--946 (2006).
\item
Ersoy, O., Gourgaud, A., Aydar, E., Chinga, G. \& Thouret, J.-C. Quantitative SEM analysis of volcanic ash surfaces: application to the 1982-83 Galunggung eruption (Indonesia). \textit{Geol. Soc. Am. Bull.} \textbf{119}, 743--752 (2007).
\item
Maria, A. \& Carey, S. Quantitative discrimination of magma fragmentation and pyroclastic transport process using the fractal spectrum technique. \textit{J. Volcanol. Geotherm. Res.} \textbf{161}, 234--246 (2007).
\item
Caballero, L., Sarocchi, D., Borselli, L. \& C\'ardenas, A. I. Particle interaction inside debris flows: Evidence through experimental data and quantitative clast shape analysis. \textit{J. Volcanol. Geotherm. Res.} \textbf{231-232}, 12-23 (2012).
\item
Liu, E.J., Cashman, K.V. \& Rust, A.C. Optimising shape analysis to quantify volcanic ash morphology. \textit{GeoResJ} \textbf{8}, 14--30 (2015).
\item
Leibrandt, S. \& Le Pennec, J. Towards fast and routine analyses of volcanic ash morphometry for eruption surveillance applications. \textit{J. Volcanol. Geotherm. Res.} \textbf{297}, 11--27. (2015).
\item
Miwa, T., Shimano, T. \& Nishimura, T. Characterization of the luminance and shape of ash particles at Sakurajima volcano, Japan, using CCD camera images. \textit{Bull. Volcanol.} \textbf{77}, 5 (2015).
\item
Fitch, E. P., Fagants, S. A., Thordarson, T. \& Hamilton, C. W. Fragmentation mechanism associated with explosive lava--water interactions in a lacustrine environment. \textit{Bull. Volcanol.} \textbf{79}, 12 (2017).
\item
Schmith, J., H\"oskuldsson, \'A \& Holm, P. M. Grain shape of basaltic ash populations: implications for fragmentation. \textit{Bull. Volcanol.} \textbf{79}, 14 (2017).
\item
Avery, M. R., Panter, K. S., Gorsevski, P. V. Distinguishing styles of explosive eruptions at Erebus, Redoubt and Taupo volcanoes using multivariate analysis of ash morphometrics. \textit{J. Volcanol. Geotherm. Res.} \textbf{332}1--13 (2017).
\item
Nurfiani, D. \& Bouvet de Maisonneuve, C. Furthering the investigation of eruption styles through quantitative shape analyses of volcanic ash particles. \textit{J. Volcanol. Geotherm. Res.} \textbf{354}, 102-114 (2017).
\item
Malvern Instruments Ltd. Morphologi G3 User Manual. (2008).
\item
Krizhevsky, A., Sutskever, I. \& Hinton, G. E. Imagenet classification with deep convolutional neural networks. \textit{Advances in Neural Information Processing Systems} \textbf{25}, 1106--1114 (2012).
\item
Simard, P. Y., Steinkraus, D. \& Platt, J. C. Best practices for convolutional neural networks applied to visual document analysis. \textit{ICDAR} \textbf{3}, 958--962 (2003).
\item
Lawrence, S., Giles, C. L., Tsoi, A. C. \& Back, A. D. Face recognition: A convolutional neural-network approach. \textit{IEEE transactions on neural networks} \textbf{8}, 98--113 (1997).
\item 
Thordarson, T. \& H\"oskuldsson, \'A. Iceland, {\it Terra Publishing}, Edinburgh, 200 p (2002).

\bibitem{Tsukui2005}
Tsukui, M., Kawanabe, Y. \& Niihori, K. 
Miyakejima Geological Map,
{\it}
Geological Survey of Japan, AIST (2005). 

\bibitem{Hasebe2001}
Hasebe, N., Fukutani, A., Sudo, M. \& Tagami, T.
Transition of eruptive style in an arc--arc collision zone: K--Ar dating of Quaternary monogenetic and polygenetic volcanoes in the Higashi-Izu region, Izu peninsula, Japan,
{\it Bull. Volcanol.,} 
\textbf{63},
377--386 (2001).

\bibitem{Koyama2010}
Koyama, M.
Geological history of Izu area,
{\it The Shizuoka Shimbun,} 
303p (2010) (In Japanese).*

\bibitem{Koyama1995}
Koyama, M., Hayakawa, Y. \& Arai, F.
Eruptive history of  the Higashi--Izu monogenetic volcano field 2: mainly on volcanoes older than 32,000 years ago,
{\it Bull. Volcanol. Soc. Japan,}
\textbf{40},
3, 191--209 (1995) (In Japanese).

\bibitem{Fujii1984}
Fujii, T., Aramaki, S., Fukuoka, T. \& Chiba, T.
Petrology of the ejecta and lavas of the 1983 eruption of Miyake-jima,
{\it Bull. Volc. Society of Japan},
\textbf{2},
(29), 266--282 (1984) (in Japanese with English abstract).

\bibitem{Hauptfleisch2012}
Hauptfleisch, U.
High-resolution palaeolimnology of Lake M\'yvatn, Iceland,
{\it PhD thesis of University of Iceland} 
\textit{}, Reykjavik (2012).

\bibitem{Hamuro1985}
Hamuro, K.
Petrology of the Higashi-Izu Monogenetic Volcano Group,
{\it Bull. the Earthquake Res. Inst., Univ. Tokyo,} 
University of Tokyo,
\textbf{60},
335--400 (1985).

\bibitem{Thorarinsson1953}
Thorarinsson, S.
The crater groups in Iceland,
{\it Bull. volcanologique,} 
\textbf{2},
1--44 (1953).

\bibitem{Yusa1970}
Yusa, Y. \& Kuroda, N.
Geology and petrology of Izu-Takatsukayama and Funabara volcano,
{\it Geoscience reports of Shizuoka Univ.,} 
\textbf{2}(1),
43--54 (1970) (In Japanese).*

\bibitem{Kikuchi2004}
Kikuchi, K. \& Takahashi, M.
Petrology for volcanic rocks produced by nearly contemporaneous eruptions of aligned volcanic centers in the Higashi-Izu Monogenetic Volcano Group, central Japan,
{\it Bull. Inst. Natural Sci., College of Humanities and Sci., Nihon Univ.,}
\textbf{39},
217--246 (2004) (in Japanese with English abstract).

\bibitem{Miyajima1990}
Miyajima, H.
Petrology of Higashi-Izu monogenetic volcano group --Implication of xenocrysts, time and spatial variation of ejecta--,
{\it Mineralogy, petrology and economic geology (Ganko),}
\textbf{85} (7) ,
315--336 (1990) (in Japanese with English abstract).

\bibitem{Hoskuldsson2010}
H\"oskuldsson, \'A., Dyhr C. \& Dolvik, T. 
Gr\ae navatnsbruni og Lax\'arhraun yngra,
{\it Haustr\'a dstefna Jardfr\ae daf\'elags \'Islands, \'Agrip erinda,}
41--44 (2010) (in Icelandic).

\bibitem{Thorarinsson1951}
Thorarinsson, S.
Lax\'arglj\'ufur and Lax\'arhraun: A Tephrochronological Study,
{\it Geografiska Annaler,} 
\textbf{33},
1--89 (1951).


\item
Aramaki, S., Hayakawa, Y., Fujii, T., Nakamura, K. \& Fukuoka, T. The October 1983 eruption of Miyakejima volcano. \textit{J. Volcanol. Geotherm. Res.} \textbf{29}, 203--229 (1986).

\bibitem{Wohletz1983}
Wohletz, K. H.
Mechanisms of hydrovolcanic pyroclast formation: particle-size, scanning electron microscopy, and experimental studies,
{\it J. Volcanol. Geotherm. Res.} 
\textit{17} (1-4),
31--63 (1983).
\item
Coltelli, M., Miraglia, L. \& Scollo, S. Characterization of shape and terminal velocity of tephra particles erupted during the 2002 eruption of Etna volcano, Italy. \textit{Bull Volcanol} \textbf{70} 1103 (2008).
\item
Fisher, R. V. \& Schmincke, H. U. Volcanic sediment transport and deposition. In: Pye, K. (ed.): Sediment transport and depositional processes, Blackwell Scientific, Oxford, 349--386 (1994).
\item
White, J. Pre-emergent construction of a lacustrine basaltic volcano, Pahvant Butte, Utah (USA). \textit{Bull. Volcanol.} \textbf{58}: 249 (1996).

\item
Chollot, F. Keras: Deep learning library for theano and tensorflow. https://keras.io (2016).
\item
Srivastava, N., Hinton, G., Krizhevsky, A., Sutskever, I. \& Salakhutdinov, R. Dropout: A simple way to prevent neural networks from overfitting. \textit{The Journal of Machine Learning Research,} \textbf{15}, 1929-1958 (2014).
\item
Glorot, X., \& Bengio, Y. Understanding the difficulty of training deep feedforward neural networks. In \textit{Proceedings of the Thirteenth International Conference on Artificial Intelligence and Statistics} 249-256 (2010).
\item
Kingma, D. P. \& Ba, J. Adam: A Method for Stochastic Optimization. arXiv:1412.6980 (2015).

\item
Lowe, D. J., Pearce N. J. G., Jorgensen, M. A., Kuehn, S. C., Tryon, C. A. \& Hayward, C. L. Correlating tephras and cryptotephras using glass compositional analyses and numerical and statistical methods: Review and evaluation. \textit{Quaternary Science Reviews} \textbf{175}, 1--44 (2017).
\item
Noguchi. R., Hino, H., Geshi, N., Otsuki, S. and Kurita, K. New classification method of volcanic ash samples using statistically determined particle types. arXiv:1712.05566 [physics.geo-ph] (2017).
\item
Anderberg, M. R. Cluster analysis for applications: probability and mathematical statistics: a series of monographs and textbook (Vol. 19), Academic press (2004).
\item
R Core Team. R: A language and environment for statistical computing. \textit{R Foundation for Statistical Computing}, Vienna Austria (2016).
\item
Hamilton, C. W., Fitch, E. P., Fagents, S. A. \& Thordarson, T. Rootless tephra stratigraphy and emplacement processes. \textit{Bull. Volcanol.} \textbf{79}, 11 (2017).

\item
B\"utter, R. Dellino, P. \& Zimanowski, B. Identifying magma--water interaction from the surface features of ash particles. \textit{Nature} \textbf{401}, 688--690 (1999).
\item
Hayward, B. W., Holzmann, M., Grenfell, H. R., Pawlowski, J. \& Triggs, C. M. Morphological distinction of molecular types in Ammonia-towards a taxonomic revision of the world's most commonly misidentified foraminifera. \textit{Marine Micropaleontology} \textbf{50}, 237--271 (2004).







\end{thebibliography}
\end{document}